\documentclass{ws-tpe}
\usepackage{multicol}
\begin{document}

\title{On Dingle's rebuttal of the special theory of relativity}

\author{Justo Pastor Lambare}

\address{Facultad de Ciencias Exactas y Naturales,Universidad Nacional de Asunci\'{o}n,Ruta Mcal. J. F. Estigarribia, Km 11 Campus de la UNA\\
San Lorenzo-Paraguay\\
\email{jupalam@gmail.com}
}


\maketitle


\begin{abstract}
In his 1972 book \emph{Science At the Crossroads}, Helbert Dingle attacked the consistency of special relativity through a fallacious argument championed by the crank community even to this day.
Dingle's affair is a curious chapter in the history of physics and, more generally, science.
We briefly review Dingle's case from a historical and didactic perspective.
\end{abstract}

\keywords{Special relativity, simultaneity, Lorentz transformation}

\begin{multicols}{2}
\section{Introduction}\label{sec:Intro}
After the publication of Einstein's 1905 seminal paper\cite{bEin07} presenting the theory of special relativity, the new theory clashed with resistance and criticism by many physicists, philosophers, and scientists.
According to Thomas Kuhn, that was a normal reaction characterizing the appearance of a paradigm shift.\cite{bKuh62}
It is remarkable, however, that the situation persisted even in the 1920s.
Among the critics at that time were eminent scientists, members of the Nobel committee, and physicists, some of whom were Nobel laureates such as Philipp Lenard and Johannes Stark.\cite{pFri22}

In more recent years, the situation of the 1920s radically changed.
However, not every anti-relativist should be readily dismissed as naive and uninformed negationist.
Some professional physicists with respectable careers occasionally hang on to deeply ingrained Newtonian prejudices and become reluctant to accept and understand the intricacies of such a counterintuitive theory despite its sound logical structure and empirical adequacy.

Two such examples are Louis Essen\cite{pEss88} and Herbert Dingle.\cite{pDin67}
Essen was an expert in the precise measurement of time and atomic clocks.
Dingle made a professional career in physics, astronomy, and philosophy of science.
He served as president of the Royal Astronomical Society from 1951 to 1953.

Both Essen and Dingle wrote books refuting relativity.\cite{bEss71,bDin72}\footnote{A review of those books can be found in Ref.~\refcite{pArm73}.}
Ironically, about the same time those books were released, experiments with atomic clocks, in which Essen was a pioneer, confirmed what these men claimed was absurd and impossible.\cite{pHK72}

Dingle's case is perhaps the most notorious one.
He wrote a popular book on relativity in 1922 and a short textbook in 1940.
For that reason, besides his academic background, Dingle was considered an expert on the subject, although he gradually became critical of it.

According to H. Chang, Dingle's anti-relativist long saga started at least in 1939, lasting until his Death in 1978 at the age of 88.\cite{pCha93}
His credentials and persistence allowed him to create a visible and lively controversy that appeared in major journals and involved physicists of the stature of Max Born.\cite{pBor63}

In section \ref{sec:DIA} we present Dingle's main argument, which he referred to as \emph{THE ARGUMENT} in his book \emph{Science At the Crossroads}.\cite{bDin72}
Section \ref{sec:TRISR} reviews the basic concepts necessary to understand why Dingle's argument is incorrect.
In section \ref{sec:M&E} we analyze Dingle's argument explaining his mistake.
\section{Dingle's inconsistency argument}\label{sec:DIA}
Dingle put forward two different arguments against relativity.
First, he criticized the twin paradox but later changed his criticism to formulate \emph{THE ARGUMENT} which we shall term here  \emph{Dingle's inconsistency argument} (DIA).

According to his chronicle in \emph{Science at the Crossroads}, in 1955, he became aware that the twin paradox implies a contradiction. Still, he later realized his mistake and branded the issue as inconclusive.

Given that he ultimately disregarded the twin paradox as his main argument and the issue is nowadays a standard exercise in introductory courses of special relativity besides having been widely discussed and explained in books, journals, and social media,\footnote{A thorough exposition of Dingle's controversy over the twin paradox can be found in Ref. \refcite{pCha93}.} we shall concentrate only on his inconsistency argument:\cite{bDin72}
\begin{quote}
\textit{According to the special theory of relativity, two similar docks, A and B, which are in uniform relative motion and in which no other differences exist of which the theory takes any account, work at different rates. The situation is therefore entirely symmetrical, from which it follows that if A works faster than B, B must work faster than A. Since this is impossible, the theory must be false.}
\end{quote}
Before entering any analysis of DIA, we shall clearly state that taken literally, DIA is correct.
Indeed, two different clocks $A$ and $B$ put together side by side, upon comparison of its readings cannot show times such that $A$ is ahead of $B$ and at the same time $B$ is ahead of $A$.
Of course, special relativity does not claim such an absurdity.

Notably, Dingle did recognize that the readings of $A$ and $B$ cannot be directly compared:\cite{bDin72}
\begin{quote}
\textit{Of course, A is not at B to allow a direct comparison, but Einstein's theory is based on a particular process for finding a clock-reading for a distant event, and it demands these values.}
\end{quote}
We shall see that the mistake in DIA consists of the incorrect understanding of the \textit{``....particular process for finding a clock-reading for a distant event...''.}
\section{Time and reality in special relativity}\label{sec:TRISR}
Before any concrete analysis of DIA, we need a clear understanding of the most fundamental implications of relativity's principles.
Special relativity relies on two premises: a) The physical equivalence of all inertial frames, and b) The light principle: light in vacuum propagates with the same speed c in all inertial frames.\footnote{A complete analysis of the foundations of relativity theory can be found in Ref. \refcite{bBro05}.}

Next, we briefly review three concepts that are crucial to refute DIA and understand the mistakes that it involves.
\subsection{The concept of time}\label{ssec:TCOT}
The definition and meaning of time are considered quite difficult philosophical problems.
From the pragmatic stance of the physicists, time is a parameter or variable necessary to express the dynamical laws of physics.
In pre-relativistic physics time was a global universal parameter as defined by Newton:\cite{SEP22}
\begin{quote}
Absolute, true, and mathematical time, from its own nature, passes equably without relation to anything external, and thus without reference to any change or way of measuring of time (e.g., the hour, day, month, or year).
\end{quote}
Instead of assuming that time is a global universal and unperturbable variable, Einstein realized that a more realistic operational definition of time was necessary.
In relativity, the time variable ``t'' is defined only within a fixed inertial frame by a mesh of stationary natural clocks that are assumed to be synchronized.\footnote{This time concept is called ``coordinate time'' to distinguish it from ``proper time'', cf. Ref. \refcite{pBac19}. There are also alternative ways to define coordinate time, see Refs. \refcite{bBon64,bboh65}.}

A natural clock is a device undergoing a cyclic process according to known physical laws.
In this way, time is defined by change according to the dynamical laws of physics and, conversely, the expression of such laws needs the existence of the time variable.
These kinds of circularity are unavoidable when defining the most fundamental concepts.\footnote{See, for instance, Feynman's discussion of force in Ref. \refcite{bFey63}.}
Only by analyzing and assessing the whole set of observable phenomena will we be able to test the appropriateness of our definition and introduce changes if necessary.
A guiding criterium is simplicity as expressed by Wheeler, Thorne, and Misner: \textit{``Time is defined so that motion looks simple''}.\cite{bMTW73}

Since according to the principle of relativity, two inertial frames $S$ and $S\,'$ are physically equivalent, within each system, through equivalent physical processes, we can define the time variables $t$ and $t'$ respectively but we do not need to assume that $t=t'$.
That shall have to be consistently decided within the theory according to its principles and empirically confirmed or falsified.
\vspace{0.2cm}
\vspace{1cm}
\begin{figurehere}
\centerline{
\includegraphics[width=8cm]{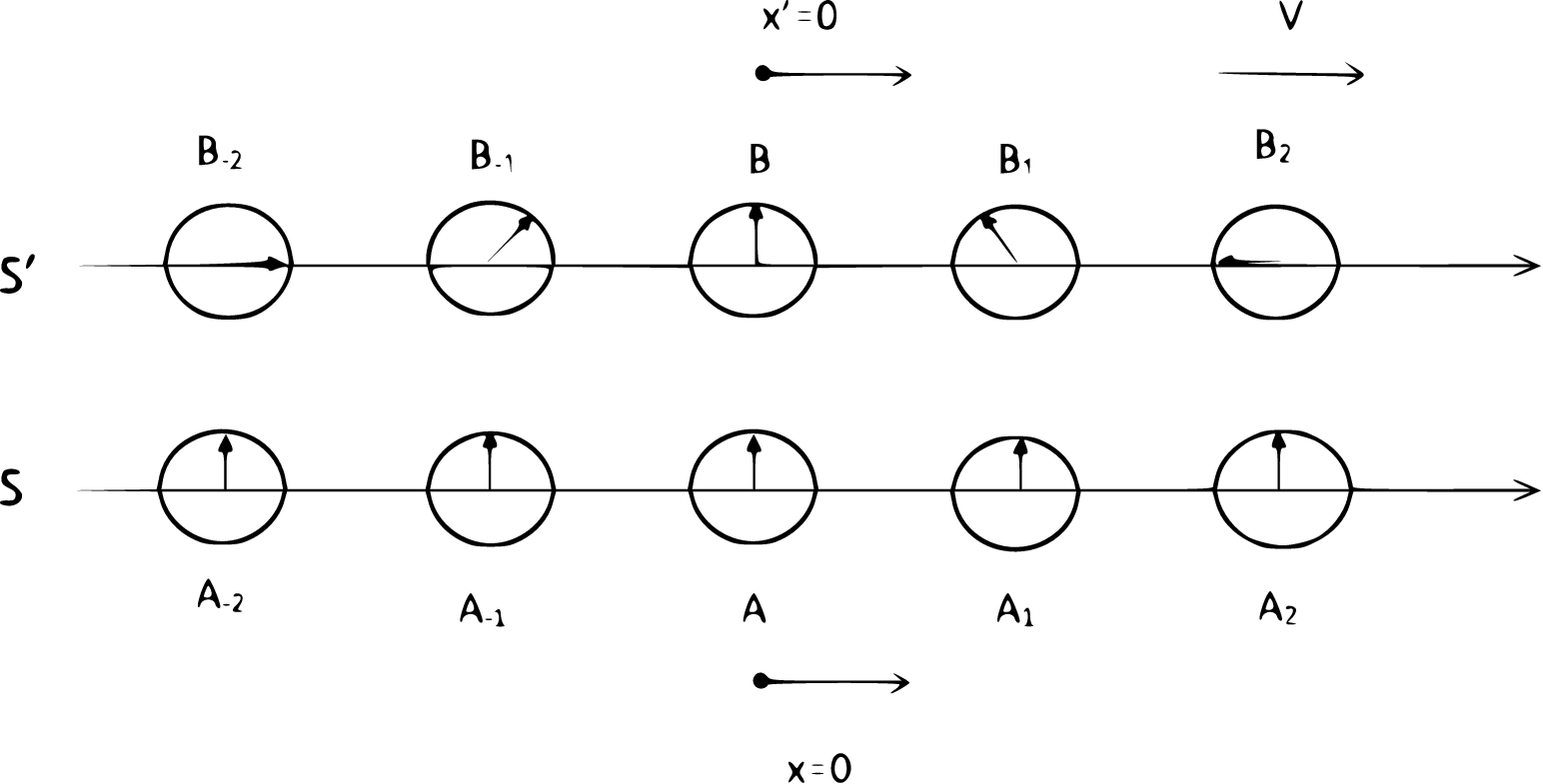} }
\caption{Mesh of synchronized clocks in $S$ and $S'$ as seen from $S$.}
\label{fig1}
\end{figurehere}
\subsection{Clock synchronization and simultaneity}\label{ssec:CSAS}
As mentioned in \ref{ssec:TCOT}, the physical definition of time within a given inertial frame requires two conditions: a) A mesh of fixed clocks distributed through space, and b) The synchronization of those clocks.

Clocks are synchronized by exchanging signals at constant and isotropic speeds.
Two distant clocks $A$ and $B$ are said to be synchronized when
\begin{equation}\label{eq:sc}
T_B=\frac{T_A + T_{A'}}{2}
\end{equation}
where $T_A$ is the time of emission of the signal at $A$, $T_B$ is the time showed by the clock $B$ at the time of reflection at $B$, and $T_{A'}$ is the time at $A$ of its return to $A$.
Since (\ref{eq:sc}) is satisfied independently of the signal speed -- provided that it is isotropic --, the signals don't necessarily have to be light beams, as is usually believed.

One immediate consequence of the light principle or more generally of the principle of finiteness of the speed of propagation of interactions,\cite{bLaL75,pLam24b} is that the simultaneity of distant events is not an absolute concept.
This can be proved through Einstein's train-embankment thought experiment.\cite{bEin05}

Let $S$ and $S\,'$ be two inertial frames in relative motion with speed $V$.
For the sake of simplicity, it is enough to consider only one spatial dimension.
Let us assume we have a net of synchronized clocks in each frame determining the time variable of each system.
Clocks $A$ and $B$ are at the origin of $S$ and $S\,'$ respectively in the so-called standard configuration so that when clock $A$ marks $0$, clock $B$ also marks $0$.
Owing to the relative character of distant simultaneity, when the clocks of $S\,'$ are seen from the perspective of $S$, they will not appear synchronized (Fig. \ref{fig1}).
Conversely, for an observer comoving with $S\,'$, the clocks of $S$ are not synchronized from his perspective.
\subsection{Reality}
According to relativity theory, objective reality is comprised of the set of all spacetime ``events'' that constitute the spacetime manifold.\footnote{To delve deeper into this concept, see Ref. \refcite{bFri83}.}

In the reference frame $S$, an event $E$ is determined by its spacetime coordinates $(t,x)$, i.e., the when and where it takes place.
Correspondingly, the same event $E$ in $S\,'$ is determined by its coordinates  $(t',x')$  with respect to that frame.

The spacetime coordinates of the same event in $S$ and $S\,'$ are related according to Lorentz transformations
\begin{equation}\label{eq:dlt}
\begin{array}{ccccc}
t' & = & \gamma t   & - & \gamma V x/c^2\\
x' & = &-\gamma Vt  & + & \gamma x
\end{array}
\end{equation}
and its inverse
\begin{equation}\label{eq:ilt}
\begin{array}{ccccc}
t & = & \gamma t'   & + & \gamma V x'/c^2\\
x & = & \gamma Vt'  & + & \gamma x'
\end{array}
\end{equation}
where $\gamma=1/\sqrt{1-\beta^2}$ with $\beta=V/c$.

The relevant point is that the time variable alone, although well-defined within each inertial frame, is insufficient to specify objective reality, only the pair $(t,x)$ determining an event does.

It is particularly relevant for understanding DIA that when in a given inertial frame $S$, for two distinct events $E_1=(t_1,x_1)$ and $E_2=(t_2,x_2)$ we have $t_1=t_2$ when $x_1\neq x_2$, it does not imply that in another inertial frame $S\,'$ we must also have $t'_1=t'_2$.
In other words, the simultaneity of distant events is not part of objective reality, only the individual events are.

The main problem of understanding relativity is accepting the last point and realizing that it does not imply any logical contradiction.
Modern physics has taught physicists to rely on rational thinking and empirical facts rather than naive everyday intuition.
\section{Analysis and explication of DIA}\label{sec:M&E}
We want to understand in every detail why DIA is false and how it is indeed possible for an observer fixed at $A$ to claim that the moving clock $B$ is lagging, while an observer fixed with $B$ claims that it is the moving clock $A$ that is lagging without implying any contradiction.
The short answer is that there is no contradiction because different clocks involving different events are being compared in each case.

We observed in section \ref{sec:DIA} that Dingle recognized the clocks cannot be directly compared, but \textit{``Einstein's theory is based on a particular process for finding a clock-reading for a distant event, and it demands these values.''}

Next, we shall see why -- provided we accept the principles of relativity -- the correct application of the particular process for finding clock readings of distant events, does not demand the contradiction that Dingle claimed; first, through a qualitative analysis and then with a numerical example given by Dingle.
\vspace{0.2cm}
\vspace{1cm}
\begin{figurehere}
\centerline{
\includegraphics[width=7cm]{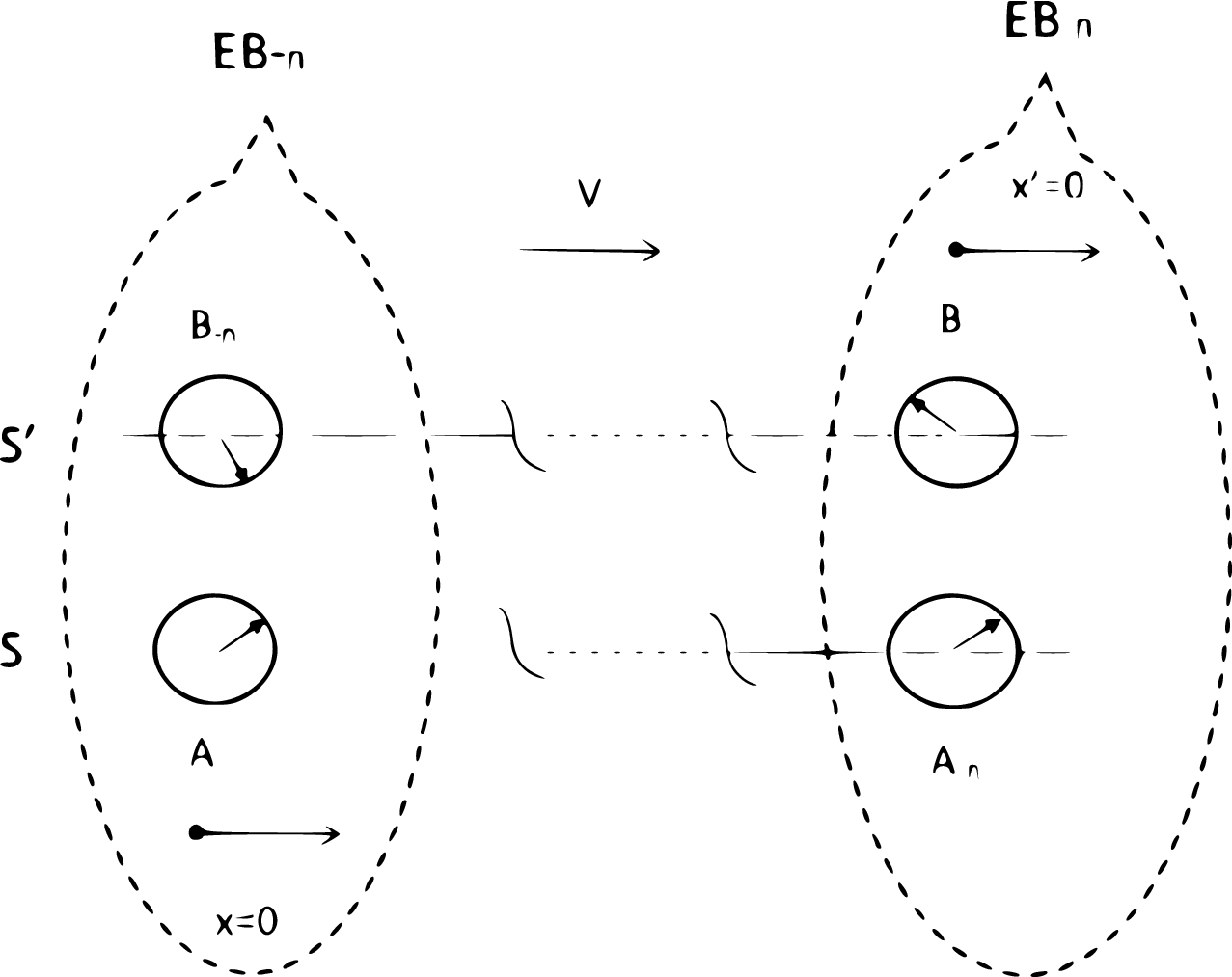} }
\caption{Comparison of clocks in relative inertial motion.}
\label{fig2}
\end{figurehere}
\subsection{Qualitative analysis}
In an article published in Nature in 1967,\cite{pDin67} Dingle used the Lorentz transformation to prove his inconsistency claim.
The paper is reproduced in an appendix to \emph{Science at the Crossroads}.
First, we give the correct solution and then comment on Dingle's mistake.

Let $S$ and $S\,'$ be two frames in the standard configuration as shown in Fig. \ref{fig1}.
Let us assume we are at clock $A$ in frame $S$ observing clock $B$ moving with $S\,'$ at speed $V$.
We assess the readings $t'_B$ of clock $B$ by comparing it with readings $t_{A_1}, t_{A_2}\ldots,t_{A_n},\ldots$  as it passes by the clocks $A_1, A_2,\ldots A_n,\ldots$ stationary in $S$.
Each comparison of readings is an event $EB_n$ happening where the clocks $B$ and $A_n$  cross each other (Fig. \ref{fig2}).

The event $EB_n$ has coordinates $(t'_B,0)$ in $S\,'$ and $(t_{A_n},x_n)$ in $S$.
The time $t'_B$ transforms according to (\ref{eq:ilt}),
\begin{eqnarray}
t_{A_n}=\gamma t'_B,\qquad \gamma>1\label{eq:tAn/t'B}
\end{eqnarray}
Let the rate-of-B/rate-of-A with $A$ fixed in $S$ be $R_{B/A}$, then from (\ref{eq:tAn/t'B}),
\begin{eqnarray}
R_{B/A} &=& t'_B/t_{A_n}=\frac{1}{\gamma}<1\label{eq:RBA}
\end{eqnarray}
When we observe clock $A$ moving from the fixed position of clock $B$ (Fig. \ref{fig2}), to compare clock readings we must consider the event $EB_{-n}$ with coordinates \begin{eqnarray}
(t'_{B_{-n}},x'_{-n})_{S'}\equiv(t_A,0)_S
\end{eqnarray}
To obtain the rate-of-A/rate-of-B with $B$ fixed we use (\ref{eq:dlt}),
\begin{eqnarray}
t'_{B_{-n}} &=& \gamma t_A,\qquad \gamma>1\label{eq:t'B-n/tA}\\
R_{A/B} &=& t_A/t'_{B_{-n}}=\frac{1}{\gamma}<1\label{eq:RAB}
\end{eqnarray}
From (\ref{eq:RBA}) and (\ref{eq:RAB}) we have,
\begin{equation}
R_{B/A}= R_{A/B}<1
\end{equation}
Thus, the correct comparison of distant clock readings shows, without contradiction, that the moving clock always lags with respect to the stationary one.
To obtain $R_{B/A}$ we compare different clocks than the ones we use to get $R_{A/B}$ and the absurdity of having the same two clocks lagging with respect to each other never happens.

Where does the mistake of Dingle and his followers lie?
To justify Dingle's solution one has to assume the absolute character of distant  simultaneity.

Indeed, if that were the case, synchronization of distant clocks has an absolute character being part of objective reality and
we must have $t_A=t_{A_n}$ and $t'_B=t'_{B_{-n}}$, thus according to (\ref{eq:RBA}) and (\ref{eq:RAB}),
\begin{equation}\label{eq:abs_RAB}
R_{B/A}=\frac{1}{R_{A/B}}
\end{equation}
Obviously, (\ref{eq:RBA}),(\ref{eq:RAB}), and (\ref{eq:abs_RAB}) are incompatible as Dingle claimed.

Furthermore, in Ref. \refcite{pDin67} he writes (\ref{eq:tAn/t'B}) and (\ref{eq:t'B-n/tA}) respectively as,
\begin{eqnarray}
t &=& \gamma t'\label{eq:tt'}\\
t'&=& \gamma t \label{eq:t't}
\end{eqnarray}
Admittedly, (\ref{eq:tt'}) and (\ref{eq:t't}) are obtained from the Lorentz transformations (\ref{eq:ilt}) and (\ref{eq:dlt}) with the standard notation.
Unfortunately, the loose application of the usual notation is misleading in Dingle's concrete problem since it induces one to assume absolute simultaneity, which is incompatible with the Lorentz transformation.
The equations (\ref{eq:tt'}) and (\ref{eq:t't}) correspond to different events, $EB_{n}$ and $EB_{-n}$ respectively, therefore we have four different variables instead of two, as the usual notation suggests.

The Lorentz transformations (\ref{eq:dlt}) and (\ref{eq:ilt}) are inverse of each other only when the same events are involved.
\subsection{Numerical example}\label{ssec:NE}
A numerical instance gains further insight into the problem of the reciprocal time lag of clocks in relative inertial motion.
In a letter to Nature in September 1962,\cite{pDin62} Dingle gave the following numerical example:
\begin{quote}
Thus, if $\sqrt{1-v^2/c^2}=1/2$ and $t=12$, Einstein's equation gives $t'=6$, and when $t'=6$  the other equation gives $t=3$. Hence, when $B$ reads 6, $A$ read both 12 and 3. That is a contradiction.
\end{quote}
When he says that $t=12$ and $t'=6$ he is using equation (\ref{eq:tt'}).
In this case, $A$ is is stationary and $B$ is in motion; $t=12$ is not reading of the clock $A$ but of the clock $A_n$ as shown in the right event of Fig. \ref{fig2}.

When says that for $t'=6$ the other equation gives $t=3$, he is using equation (\ref{eq:t't}) that corresponds to actual reading of $A$ corresponding to the left event of Fig. \ref{fig2}.

Thus, the readings $t=12$ and $t=3$ correspond to different clocks and events and there is no contradiction.

It is worth noticing the two events in $S\,'$ have the same value of $t'=6$, which means those events are simultaneous in $S\,'$ then, according to relativity, they cannot be simultaneous in $S$ as revealed by the time components $t=12$ and $t=3$ respectively.
Once more we can see that the incompatibility arises only when we assume absolute simultaneity requiring that all clocks in $S$ also read only one value of $t$ irrespective the reference frame.
\section{Conclusions}\label{sec:Con}
Dingle's inconsistency argument arises from the incorrect application of formalism.
The mistake can be interpreted as a consequence of implicitly assuming the absolute character of distant simultaneity.

Absolute simultaneity implies the rejection of the light principle according to the implication,
\begin{eqnarray}
\text{Light speed invariance}\implies \text{Relative simultaneity}\nonumber
\end{eqnarray}
Dingle's anti-relativist program is a non sequitur since one cannot prove the inconsistency of a theory by inferring consequences that arise from the negation of its axioms.

Although the abandonment of absolute simultaneity is puzzling to our limited human experience, it is a consequence of excluding the possibility of action at a distance which naturally excludes instantaneous distant happenings from objective reality.

Science, particularly physics, has taught us to rely on rational thinking and empirical evidence rather than naive human intuition.
\section{Final didactic observations}
Dingle's inconsistency argument can be used as a homework project for students to analyze, propose, and discuss solutions before presenting a definitive answer by the teacher.

The exercise presents an excellent opportunity to explain the puzzling consequences of realizing that simultaneity does not have to be absolute without implying logical contradictions and the importance of using an adequate notation when applying mathematical formulas.

The historical fact that Dingle never accepted the explications given by professors and experts such as the Nobel laureate Max Born and that such explanations remain easily accessible through the pages of journals such as Nature,\cite{pBor63} confronts the teacher with the task of wondering what went wrong and how it could have been done better.
\bibliographystyle{unsrt}
\bibliography{zRelativity}

\end{multicols}
\end{document}